\documentclass[
 reprint,
 superscriptaddress,
 amsmath,
 amssymb,
 aps,
]{revtex4-2}

\usepackage[T1]{fontenc}
\usepackage[utf8]{inputenc}
\usepackage[english]{babel}
\usepackage{graphicx}
\usepackage{bm}
\usepackage{braket}
\usepackage{xcolor}
\usepackage{microtype}
\usepackage{tikz}
\usepackage{subcaption}
\usepackage{hyperref}
\hypersetup{colorlinks=true,linkcolor=blue,citecolor=blue,urlcolor=blue}
\usetikzlibrary{positioning,calc}
\usepackage{amsthm}
\usepackage{dsfont}

\begin{document}

\title{Complete Hypergraph States as a Resource for Robust Quantum Metrology}

\author{Hoe Kin Mooi}
\email{hoekin001@e.ntu.edu.sg}
\affiliation{Quantum Science and Engineering Centre, Nanyang Technological University, Singapore 639798, Singapore}

\author{Leong-Chuan Kwek}
\email{kwekleongchuan@nus.edu.sg}
\affiliation{Quantum Science and Engineering Centre, Nanyang Technological University, Singapore 639798, Singapore}
\affiliation{Centre for Quantum Technologies, National University of Singapore, Singapore 117543, Singapore}
\affiliation{National Institute of Education, Nanyang Technological University, 1 Nanyang Walk, Singapore 637616, Singapore}

\date{\today}

\begin{abstract}
Hypergraph states extend graph states by incorporating many-body interactions via multi-controlled-$Z$ gates, but their metrological advantages and design principles remain elusive. We develop an analytical expression for the quantum Fisher information (QFI) of complete weighted hypergraphs with respect to the collective-spin generators $J_x$ and $J_y$, reducing to three finite binomial averages requiring $O(n|K|)$ arithmetic for a set of hyperedge orders $K$. At the entangling phase $\phi=\pi$, order-optimized $J_x$ sensing approaches the Heisenberg limit (HL), while $J_y$ oscillates between the HL and half the HL at large total number of qubits, $n$. We derive the quadratic and linear coefficients of the QFI under erasure and dephasing noise, showing that complete hypergraph states retain Heisenberg scaling for finite erased particles and dephasing probability $p<0.5$ in the asymptotic system size limit. For $J_y$ sensing, computational basis measurement, or, more simply, using Hamming weight detection, is sufficient to saturate the QFI under noiseless conditions and retain Heisenberg scaling under both noises. These results suggest that complete hypergraphs are a robust resource for practical metrological sensing tasks.
\end{abstract}

\maketitle

\section{Introduction}

Quantum metrology exploits uniquely quantum features to enhance the precision of parameter estimation beyond what is achievable with classical resources \cite{giovannetti_quantum-enhanced_2004,giovannetti_quantum_2006,pezze_quantum_2018}. Quantum Fisher information (QFI) quantifies the ultimate sensitivity allowed by quantum mechanics via the quantum Cram\'er--Rao bound \cite{braunstein_statistical_1994,hyllus_fisher_2012,toth_quantum_2014}. For $n$ independent probes, the achievable precision is typically bounded by the standard quantum limit (SQL), which scales as \(1/\sqrt{n}\). By using entangled states, however, the precision can in principle reach the Heisenberg limit (HL), which scales as \(1/n\). In the quantum Fisher information \(F_Q\) framework, these two regimes correspond to the scalings \(F_Q \sim n\) and \(F_Q \sim n^2\), respectively. Multipartite entangled states are a powerful resource for quantum metrology. In particular, the Greenberger--Horne--Zeilinger (GHZ) state achieves the Heisenberg limit in the noiseless regime, but is more fragile under decoherence and particle-loss noise \cite{giovannetti_quantum_2006,ma_quantum_2011,huang_quantum_2015}. 

Graph states, a class of multipartite entangled states generated through pairwise controlled-phase interactions, have broad applications in quantum computation, communication, and metrology due to their stabilizer formalism \cite{hein_multiparty_2004,guhne_entanglement_2014,shettell_graph_2020}. In particular, prior work in graph state metrology has shown that they can achieve a higher QFI than the SQL limit \cite{shettell_graph_2020}. Hypergraph states generalize graph states, where hyperedges connect more than two vertices and generate higher-order phase correlations \cite{rossi_quantum_2013,gachechiladze_changing_2019,qu_encoding_2013}. These richer entanglement correlations in hypergraph states may offer new metrological advantages, especially in regimes where higher-order correlations interact nontrivially with noise. Complete graphs, or equivalently complete 2-uniform hypergraph states, are symmetric and permutation-invariant, can achieve the HL QFI under collective $J_y$ sensing, but retain QFI below the SQL limit once one particle is lost \cite{shettell_graph_2020}. Prior work on complete hypergraph states has suggested a possible approach: based on their nonlocality properties, certain 3-uniform hypergraph states were argued to be compatible with Heisenberg-limited metrology and more robust to noise and particle loss than GHZ states \cite{gachechiladze_extreme_2016}. Therefore, a natural question is how the metrological performance of complete $ k$-uniform hypergraph states is affected by noise.

Using the permutational symmetry of complete hypergraph states, we derive the analytical formula for the noiseless QFI under collective spin $J_x$ and $J_y$ expressed by finite binomial formulas rather than the exponentially large state vectors. In the large-$n$ limit, we show that the noiseless QFI of $J_x$ approaches HL, while the latter holds true for system sizes of $n=3k_m-4$, where the order $k_m=2^m+2$. We further show that, for fixed $k\ge3$, these states retain a strictly positive $n^2$ QFI coefficient after any fixed finite number of pre-encoding particle erasures and under independent pre-encoding phase flips with any fixed probability $p<0.5$. Furthermore, local Z-basis measurements are sufficient to retain Heisenberg scaling under both noises in $J_y$ sensing. These results suggest that complete hypergraphs are a promising candidate for robust quantum metrology.

\begin{figure}
    \centering

    \begin{subfigure}[c]{0.4\columnwidth}
        \centering
        \resizebox{\linewidth}{!}{%
            \begin{tikzpicture}[scale=1.4, line cap=round, line join=round]

\fill[white!15] (-0.8,-0.8) rectangle (3.8,3.8);

\coordinate (v1) at (0,3);
\coordinate (v2) at (3,3);
\coordinate (v3) at (3,0);
\coordinate (v4) at (0,0);

\draw[purple!70, line width=1.5pt, dotted] (v1) -- (v2);
\draw[purple!70, line width=1.5pt, dotted] (v1) -- (v3);
\draw[purple!70, line width=1.5pt, dotted] (v1) -- (v4);
\draw[purple!70, line width=1.5pt, dotted] (v2) -- (v3);
\draw[purple!70, line width=1.5pt, dotted] (v2) -- (v4);
\draw[purple!70, line width=1.5pt, dotted] (v3) -- (v4);

\fill (v1) circle (2.2pt);
\fill (v2) circle (2.2pt);
\fill (v3) circle (2.2pt);
\fill (v4) circle (2.2pt);


\end{tikzpicture}%
        }
        \caption{}
        \label{fig:complete-graph}
    \end{subfigure}
    \hfill
    \begin{subfigure}[c]{0.4\columnwidth}
        \centering
        \resizebox{\linewidth}{!}{%
            \begin{tikzpicture}[scale=1.4, line cap=round, line join=round]

\fill[white!15] (-0.8,-0.8) rectangle (3.8,3.8);

\coordinate (v1) at (0,3);
\coordinate (v2) at (3,3);
\coordinate (v3) at (3,0);
\coordinate (v4) at (0,0);

\coordinate (v1o) at ($(v1)+(-0.24, 0.24)$);
\coordinate (v2o) at ($(v2)+( 0.24, 0.24)$);
\coordinate (v3o) at ($(v3)+( 0.24,-0.24)$);
\coordinate (v4o) at ($(v4)+(-0.24,-0.24)$);

\draw[blue, line width=1.2pt]
plot[smooth cycle, tension=0.5] coordinates {
    (v1o) (1.2,3.40) (v2o) (3.50,1.6) (v3o) (1.2,1.2)
};

\draw[red!80!brown, line width=1.2pt]
plot[smooth cycle, tension=0.5] coordinates {
    (v1o) (1.6,3.50) (v2o) (2.00,1.45) (v4o) (-0.35,1.4)
};

\draw[orange!80!yellow, line width=1.2pt]
plot[smooth cycle, tension=0.5] coordinates {
    (v1o) (-0.45,1.45) (v4o) (1.6,-0.55) (v3o) (2.00,1.7)
};

\draw[green!50!black, line width=1.2pt]
plot[smooth cycle, tension=0.5] coordinates {
    (v2o) (3.40,1.45) (v3o) (1.45,-0.40) (v4o) (1.25,1.65)
};

\draw[purple!70, line width=1.5pt, dotted] (v1) -- (v2);
\draw[purple!70, line width=1.5pt, dotted] (v1) -- (v3);
\draw[purple!70, line width=1.5pt, dotted] (v1) -- (v4);
\draw[purple!70, line width=1.5pt, dotted] (v2) -- (v3);
\draw[purple!70, line width=1.5pt, dotted] (v2) -- (v4);
\draw[purple!70, line width=1.5pt, dotted] (v3) -- (v4);

\fill (v1) circle (2.2pt);
\fill (v2) circle (2.2pt);
\fill (v3) circle (2.2pt);
\fill (v4) circle (2.2pt);


\end{tikzpicture}%
        }
        \caption{}
        \label{fig:complete-2-3-uniform}
    \end{subfigure}

    \caption{Complete hypergraphs on four vertices.
    (a) The complete $2$-uniform hypergraph $K_4^{(2)}$, equivalent
    to the complete graph $K_4$. (b) The complete $(2,3)$-uniform
    hypergraph $K_4^{(2,3)}$.}
    \label{fig:complete-graph-hypergraph}
\end{figure}
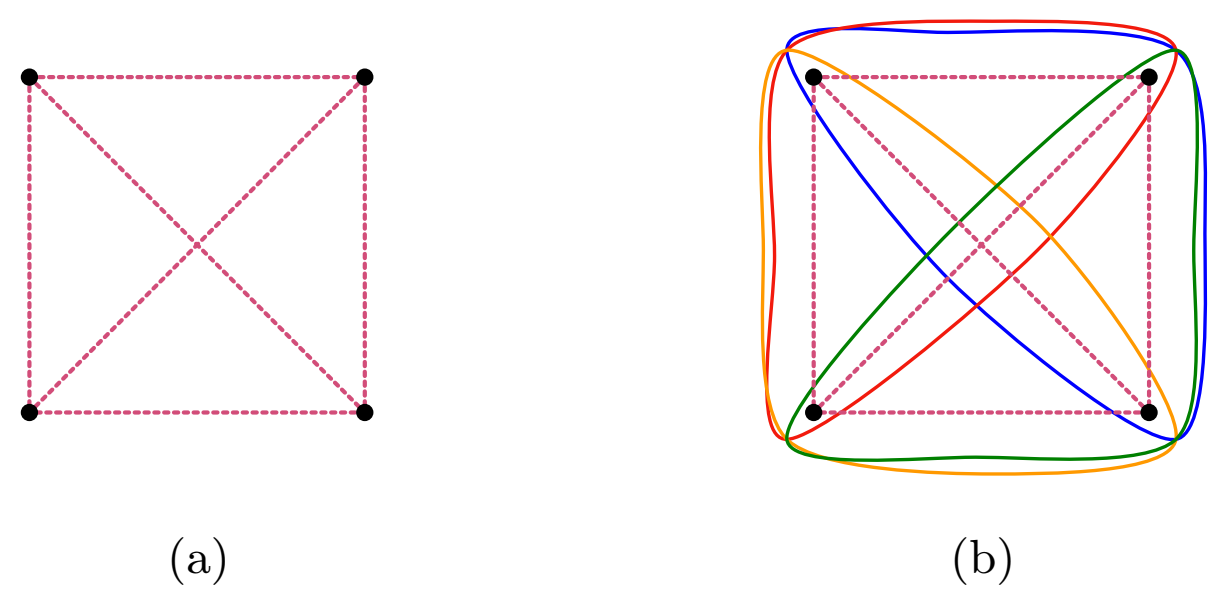

\section{Analytic hypergraph-state design}

For a set of complete hyperedge orders $K\subseteq\{2,\ldots,n\}$, define
\begin{equation}
    \begin{aligned}
        \ket{H_n(\phi;K)}&=2^{-n/2}\sum_x e^{i\Theta(|x|)}\ket{x},\\
        \Theta(w)&=\phi\sum_{k\in K}\binom wk .
    \end{aligned}
    \label{eq:state}
\end{equation}
$\phi$ controls the shared gate angle and $K$ selects the ordering of the complete hypergraphs. For single-order complete hypergraphs, $\ket{H_{n,k}(\phi)}=\ket{H_n(\phi;\{k\})}$. The parameter is encoded by
\begin{equation}
    \rho_\theta=e^{-i\theta J_\mu}\rho\,e^{i\theta J_\mu},
    \label{eq:parameter_encoding}
\end{equation}
where
\begin{equation}
    J_\mu=\frac{1}{2}\sum_{i=1}^{n}\sigma_\mu^{(i)},
    \qquad
    \mu\in\{x,y\}.
    \label{eq:collective_generator}
\end{equation}
Here, $\sigma_x^{(i)}=X_i$ and $\sigma_y^{(i)}=Y_i$ denote the Pauli-$X$ and Pauli-$Y$ operators acting on the $i$th qubit, respectively. Fixing $J_x$ or $J_y$ describes a common sensing task, but it is not necessarily the optimal Hamiltonian for every probe. When the collective axis rotation is free, the optimal generator can instead be obtained from the largest eigenvalue of the collective-spin covariance matrix \cite{hyllus_not_2010}. 

For a pure probe state $\ket{\psi}$, the QFI is four times the variance of the encoding generator,
\begin{equation}
    F_Q[\ket{\psi},J_\mu]=4\operatorname{Var}_{\psi}(J_\mu)=4\left(\langle J_\mu^2\rangle-\langle J_\mu\rangle^2\right).
    \label{eq:qfi_pure}
\end{equation} 
For a mixed state with spectral decomposition
\begin{equation}
    \rho=\sum_a \lambda_a \ket{a}\bra{a},
    \label{eq:rho_spectral}
\end{equation}
the QFI is
\begin{equation}
    F_Q[\rho,J_\mu]=2\sum_{\substack{a,b\\\lambda_a+\lambda_b>0}}\frac{(\lambda_a-\lambda_b)^2}{\lambda_a+\lambda_b}\left|\langle a|J_\mu|b\rangle\right|^2.
    \label{eq:qfi_mixed}
\end{equation}
The restriction $\lambda_a+\lambda_b>0$ excludes terms with vanishing denominators.

The exact closed-form QFI expressions are derived from the single- and two-body correlators. Define

\begin{equation}
    \begin{aligned}
        \Delta_r^{(1)}&=\phi\sum_{k\in K}\binom r{k-1},\\
        \Delta_u^{(2)}&=\phi\sum_{k\in K}\left[2\binom u{k-1}+\binom u{k-2}\right],
    \end{aligned}
\end{equation}
and put
\begin{equation}
    \begin{aligned}
        C_1&=2^{1-n}\sum_{r=0}^{n-1}\binom{n-1}{r}\cos\Delta_r^{(1)},
        \\
        S_1&=2^{1-n}\sum_{r=0}^{n-1}\binom{n-1}{r}\sin\Delta_r^{(1)},
        \\
        C_2&=2^{1-n}\sum_{u=0}^{n-2}\binom{n-2}{u}\cos\Delta_u^{(2)}.
    \end{aligned}
    \label{eq:noiseless-averages}
\end{equation}

For every $n\ge2$, every $K\subseteq \{2,\ldots,n\}$, every real $\phi$, the pure state QFIs of Eq.\eqref{eq:state} under collective generators of $J_x$ and $J_y$ are
\begin{equation}
    \begin{aligned}
        F_Q^{(x)}&=n+n(n-1)(\tfrac12+C_2)-n^2C_1^2,\\
        F_Q^{(y)}&=n+n(n-1)(\tfrac12-C_2)-n^2S_1^2 .
    \end{aligned}
    \label{eq:qfi}
\end{equation}
where $F_Q^{(x)}$ and $F_Q^{(y)}$ denotes the QFI under $J_x$ and $J_y$ respectively. Equation\eqref{eq:qfi} depends on phase, order composition, and the generator is carried by three one-dimensional averages. They scale as $O(n|K|)$ arithmetic rather than a $2^n$ state vector. This permutation-invariant state can also be expanded in the symmetric Dicke basis \cite{dicke_coherence_1954,stocktonCharacterizingEntanglementSymmetric2003}. The normalized $n$-qubit Dicke state with $w$ excitations is 
\begin{equation}
    \ket{D_n^w}=\binom{n}{w}^{-1/2}\sum_{\substack{x\in\{0,1\}^n\\ |x|=w}}\ket{x},
    \qquad
    w=0,\ldots,n,
    \label{eq:dicke-basis}
\end{equation}
where $|x|=\sum_{i=1}^{n}x_i$ denotes the Hamming weight of the bit string $x$.  The states $\{\ket{D_n^w}\}_{w=0}^{n}$ form an orthonormal basis of the fully symmetric $n$-qubit subspace. For even $n$, $|D_n^{n/2}\rangle=|j=n/2,m=0\rangle$ has $F_Q[J_x]=F_Q[J_y]=n(n+2)/2$ and is the spin analog of the twin-Fock resource in Holland--Burnett interferometry \cite{holland_interferometric_1993,toth_quantum_2014}; Dicke-like multipartite entanglement has also been certified experimentally \cite{luckeDetectingMultiparticleEntanglement2014}. The complete hypergraph state in the Dicke basis is
\begin{equation}
    \ket{H_n(\phi;K)}=\sum_{w=0}^n\beta_w^{(n)}|D_n^w\rangle,
    \label{eq:dicke_H}
\end{equation}
where $\beta_w^{(n)}={2^{-n/2}}\sqrt{\binom{n}{w}}e^{i\Theta(w)}$.

\begin{figure}
    \centering
    \includegraphics[width=\linewidth]{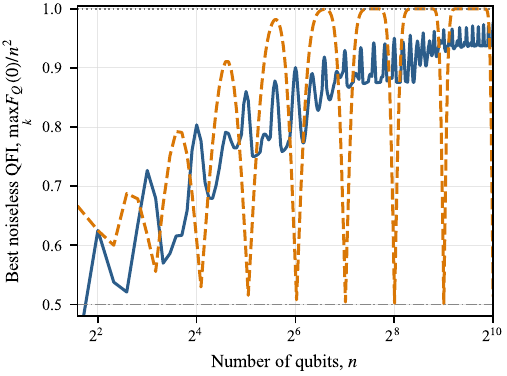}
    \caption{Exact $\pi$-phase single-order envelopes.  At every integer $n=3,\ldots,2^{10}$, the plotted value is $\max_{3\leq k\leq n}F_Q(0)/n^2$ at $\phi=\pi$. Blue and orange denote $J_x$ and $J_y$, respectively.  The black dotted line marks the Heisenberg value $1$, while the black dash-dotted line marks the asymptotic $F_Q^{(y)}$ value at $1/2$ in Eq.~\eqref{eq:noiseless-scaling}.}
    \label{fig:noiseless-envelopes}
\end{figure}

To certify the Heisenberg scaling of a state, we introduce the Heisenberg coefficients, which are the Fisher information over the square of the number of particles $n$ in the limit $n\to\infty$. Let $s_2(k)$ be the number of ones in binary $k$ and define 
\begin{equation}
    a_k=2^{-s_2(k-2)},\qquad
    b_k=1-2^{1-s_2(k-1)}.
    \label{eq:binary-coefficients-main}
\end{equation}
The noiseless Heisenberg coefficient for every $k\ge2$ and $\phi=\pi$ is
\begin{equation}
    \begin{aligned}
        \lim_{n\to\infty}\frac{F_Q^{(x)}}{n^2}&:=c_{0,x}^{(k)}=1-a_k-b_k^2,\\
        \lim_{n\to\infty}\frac{F_Q^{(y)}}{n^2}&:=c_{0,y}^{(k)}=a_k.
    \end{aligned}
    \label{eq:noiseless-fixed-coefficients}
\end{equation}

For each positive integer $m\in\mathbb{N}=\{1,2,3,\ldots\}$, choose $k_x(m)=2^m+1$ and $k_y(m)=2^m+2$ then yields 
\begin{equation}
    \begin{aligned}
        \lim_{n\to\infty}\frac{F_Q^{(x)}}{n^2}&=1-2^{-m},\\
        \lim_{n\to\infty}\frac{F_Q^{(y)}}{n^2}&=\frac{1}{2}.
    \end{aligned}
    \label{eq:noiseless-scaling}
\end{equation}
The $J_x$ envelope therefore approaches the Heisenberg limit as the order $k$ increases with $m$, but the $J_y$ envelope asymptotically reaches only half of the HL. By allowing the order $k$ to be dependent on the particle number $n$, then when $n(m)=3\cdot2^m+2$ and $k_y(m)=2^m+2$,
\begin{equation}
    \lim_{m\to\infty}
    \frac{F_Q^{(y)}(n(m),k_y(m),\pi)}{n(m)^2}=1.
\end{equation}
Thus, as $n$ and $k$ grow together, the $J_y$ envelope contains resonant points whose normalized QFI approaches the HL. For $k_m=2^m+2$, the same order is optimal approximately over the range $2^{m+1}+2\leq n\leq2^{m+2}$, with the near-HL resonance at $n_m=3\cdot2^m+2$. Accordingly, Fig.~\ref{fig:noiseless-envelopes} shows resonant peaks at $(n,k)=(386, 130)$ and $(770, 258)$. Between the peak resonances, the boundary size $n=2^m+1$ forms troughs at which no higher-order $k\geq3$ has a leading normalized QFI coefficient exceeding $1/2$.

\section{Exact QFI under erasure noise}

Complete weighted hypergraph states are permutation invariant whenever the gate phase is uniform within each hyperedge order. Consequently, the reduced state depends only on the number of erased particles, not on their identities. Let $e$ particles be erased and $m=n-e$ remain. The Dicke-state bipartition identity is
\begin{equation}
    \ket{D_n^w}=\sum_{t=\max(0,w-m)}^{\min(e,w)}\sqrt{\frac{\binom{m}{w-t}\binom{e}{t}}{\binom{n}{w}}}\,\ket{D_m^{\,w-t}}\otimes \ket{D_e^{\,t}} .
    \label{eq:dicke_split}
\end{equation}
Tracing out the erased subsystem therefore gives
\begin{equation}
    \rho_m^{(e)}=\sum_{r,s=0}^{m}M_{rs}^{(e)}\,\ket{D_m^r}\bra{D_m^s},
    \label{eq:reduced_erasure_state}
\end{equation}
with
\begin{equation}
    \begin{aligned}
        M_{rs}^{(e)}&=\sqrt{\binom{m}{r}\binom{m}{s}}\sum_{t=0}^{e}\binom{e}{t}\frac{\beta_{r+t}^{(n)}\beta_{s+t}^{*(n)}}{\sqrt{\binom{n}{r+t}\binom{n}{s+t}}}\\
        &=
        \frac{\sqrt{\binom{m}{r}\binom{m}{s}}}{2^n}\sum_{t=0}^{e}\binom{e}{t}e^{i[\Theta(r+t)-\Theta(s+t)]}.
    \label{eq:reduced_erasure_coeff}
    \end{aligned}
\end{equation}

Thus the exact QFI of the retained state is obtained by diagonalizing only the $(m+1)\times(m+1)$ matrix $M^{(e)}$ and applying Eq.~\eqref{eq:qfi_mixed} with the generator $J_\mu$.

The reduction also makes the conditional action of particle loss explicit. For a single $k$-uniform layer, Vandermonde's identity gives
\begin{equation}
    \binom{r+t}{k}-\binom{r}{k}=\sum_{q=1}^{k}\binom{t}{q}\binom{r}{k-q}.
    \label{eq:channel}
\end{equation}
For $k=2$, the difference is $tr+\binom{t}{2}$. Conditional on the erased subsystem having Hamming weight $t$, the second term contributes only a $t$-dependent global phase, whereas the first produces a collective rotation of the retained particles about $z$. Averaging over the erased weight gives
\begin{equation}
    \rho_{m,2}^{(e)}=2^{-e}\sum_{t=0}^{e}\binom{e}{t}e^{i\phi t\hat w}\ket{H_{m,2}(\phi)}\bra{H_{m,2}(\phi)}e^{-i\phi t\hat w},
    \label{eq:pairwise-channel}
\end{equation}
where $\hat w$ is the retained excitation-number operator. Since $\hat w=m/2-J_z^{(m)}$, each conditional unitary is, up to a global phase, a collective $z$ rotation. For $k\geq3$ and $t>0$, by contrast, Eq.~\eqref{eq:channel} generically produces a nonlinear polynomial of degree $k-1$ in the retained weight. Hyperedge order therefore changes the conditional channel induced by loss, not merely the noiseless QFI.

To obtain the erasure Heisenberg coefficient, $c_{e,\mu}^{(k)}$, at $\phi=\pi$, let $l_2(k) = \lfloor \log_2 k \rfloor+1$ be the number of binary digits needed to represent $k$, define
\begin{equation}
    \begin{aligned}
        L&=2^{l_2(k)},&
        s_r&=(-1)^{\binom rk},\\
        a_q&=\frac1L\sum_{r=0}^{L-1}s_r e^{-2\pi i qr/L},
    \end{aligned}
    \label{eq:fourier-amplitudes-main}
\end{equation}
and $\theta_q=2\pi q/L$.  Fourier inversion gives the exact identity
\begin{equation}
    \ket{H_{n,k}(\pi)}=\sum_{q=0}^{L-1}a_q\ket{\theta_q}^{\otimes n},\qquad
    \ket{\theta_q}=\frac{\ket{0}+e^{i\theta_q}\ket{1}}{\sqrt2}.
    \label{eq:fourier-branches-main}
\end{equation}
After $e$ erasures, the corresponding Fourier branch density matrix is
\begin{equation}
    (\sigma_e)_{qq'}=a_qa_{q'}^*\left[\frac{1+e^{i(\theta_q-\theta_{q'})}}2\right]^e.
    \label{eq:sigma-e-main}
\end{equation}
It is well noted that the discrete Fourier basis in Eq.\eqref{eq:fourier-branches-main} is not orthogonal, where 
\begin{equation}
    \left|\braket{\theta_{q'}|\theta_q}\right|^n = \left|\cos\frac{\theta_q \theta_{q'}}{2}\right|^n.
    \label{eq:non-orthogonal-branch}
\end{equation}
Nevertheless, the magnitude of the cosine term is less than 1 and therefore orthogonalized in the large limit of $n$. The normalized collective generators converge to $h_x(q)=\cos\theta_q/2$ and $h_y(q)=\sin\theta_q/2$. Because the branch matrix has dimension $L\le2k$, its spectral QFI can be evaluated in $O(k^3)$ operations, independent of the retained system size $m$. On the other hand, the exact finite-$m$ calculation using the Dicke density matrix in Eq.\eqref{eq:reduced_erasure_state} requires $O(m^3)$ evaluation. Therefore, for fixed $k$ and $e$, the Fourier-branch representation directly yields the Heisenberg coefficient $c_{e,\mu}^{(k)}$.

For $\phi=\pi$, every fixed $k\geq3$, every fixed integer $e\geq0$, and each $\mu\in\{x,y\}$, there exist finite constants $C,s$ and a rate $0\leq\Gamma<1$. With $m=n-e$,
\begin{equation}
    \begin{aligned}
        F_Q(\rho_m^{(e)},J_\mu)&=c_{e,\mu}^{(k)}m^2+d_\mu^{(k)}m+r_{e,\mu}^{(k)}(m),\\\\
        c_{e,\mu}^{(k)}&=F_Q(\sigma_e,h_\mu)>0,\\
        d_x^{(k)}&=a_k,\\
        d_y^{(k)}&=1-a_k,
    \end{aligned}
    \label{eq:erasure-expansion-main}
\end{equation}
with the remainder $|r_{e,\mu}^{(k)}(m)|\leq Cm^s\Gamma^m$. Since $F_Q(\sigma_e,h_\mu)=0$ if and only if $[\sigma_e,h_\mu]=0$, the strict positivity of the erasure Heisenberg coefficient $c_{e,\mu}^{(k)}$ is proven by the nonzero commutators using pairs of odd Fourier branches $q$ for $k>3$, while the $k=3$ case is evaluated separately.

Therefore, for every $k\ge3$ and a finite number of erasures $e$, the complete hypergraph state retains the Heisenberg scaling of the state under both $J_x$ and $J_y$ encodings. In contrast, the $k=2$ complete graph remains at SQL scaling regardless of the number of erased particles under $J_x$, whereas under $J_y$ encoding, a single erasure destroys the noiseless Heisenberg scaling, with the QFI falling below the SQL.

\begin{figure}
    \centering
    \includegraphics[width=\linewidth]{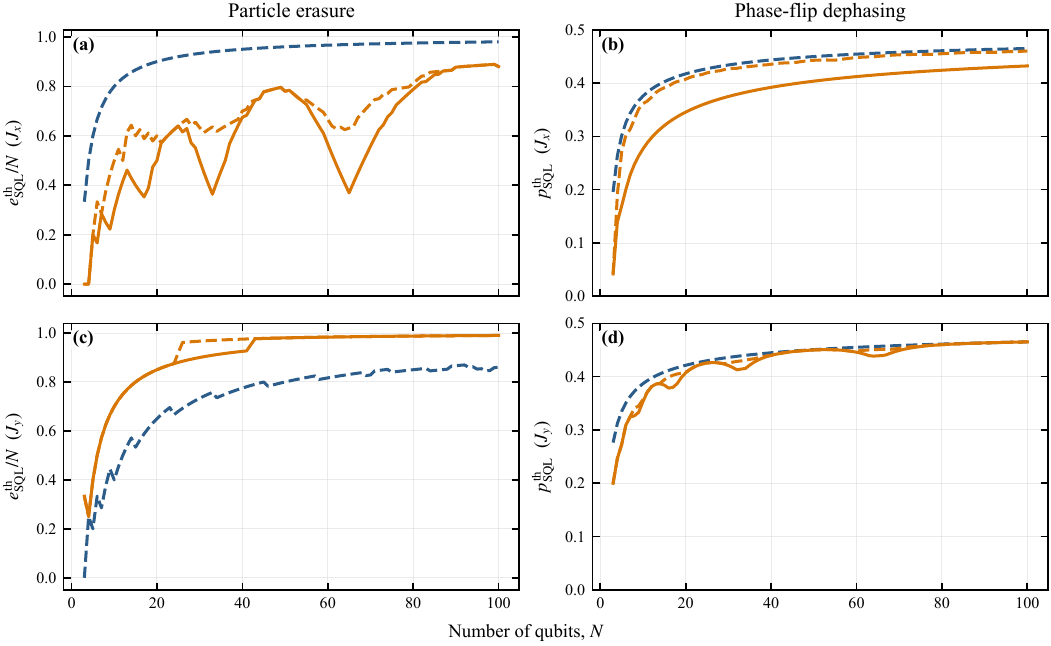}
    \caption{Noise thresholds for collective generator $J_x$ (Top) and $J_y$ (Bottom). Left: Erasure threshold from Eqs.~\eqref{eq:dicke_split}--\eqref{eq:reduced_erasure_coeff}. Right: iid dephasing threshold from Eq.~\eqref{eq:dephasing-fast}. We compare weighted complete graphs (dotted blue) with weighted (dotted orange) and $\pi$-phase complete $k$-uniform hypergraphs (solid orange). For each value of $n\in\{3,\ldots,100\}$, the order $k\geq3$ of the complete hypergraph is optimized for the largest threshold possible, while the weighted complete graph that has order $k=2$ is optimized over the shared entangling phase $\phi\in[0,\pi]$ in 721 points.  We exclude the $\pi$-complete graph because its displayed erasure threshold is zero, while its $J_y$ dephasing curve coincides with the weighted-graph envelope.}
    \label{fig:noise-thresholds}
\end{figure}

\begin{figure}
    \centering
    \includegraphics[width=\linewidth]{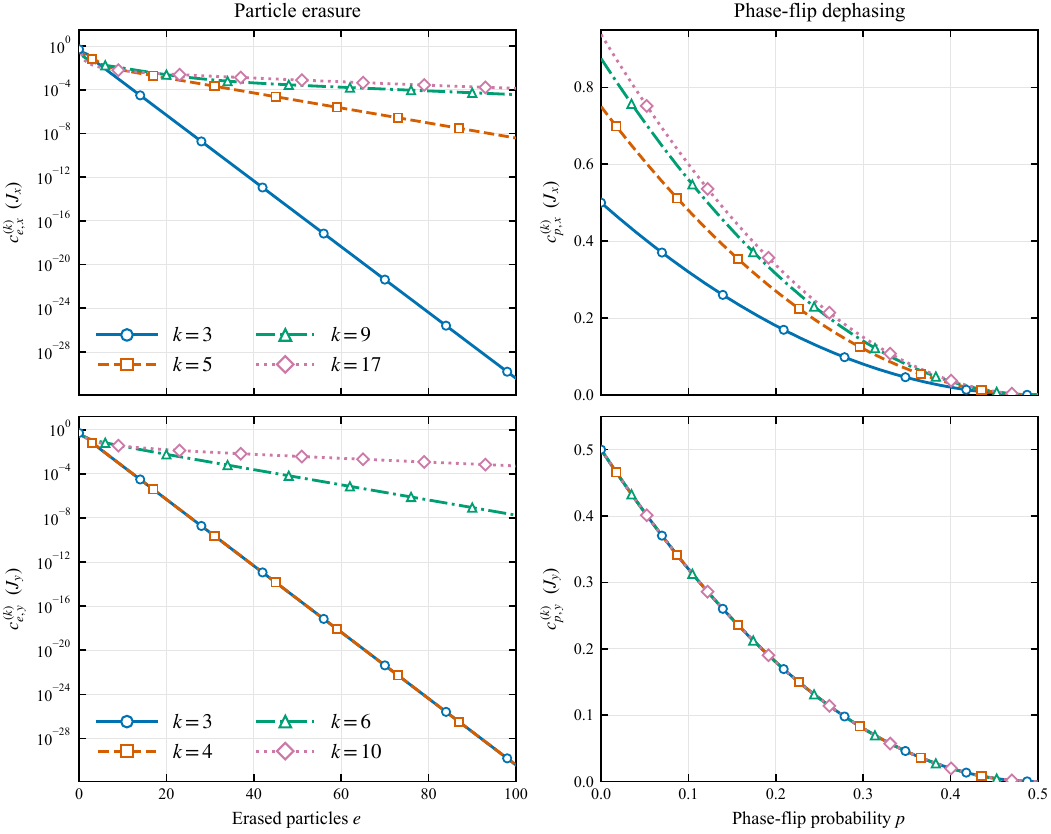}
    \caption{Asymptotic noise coefficients for collective generator $J_x$ (Top) and $J_y$ (Bottom). Left: Asymptotic Heisenberg coefficients under particle erasure from Eq. \eqref{eq:erasure-expansion-main}. Right: Asymptotic Heisenberg coefficients under phase-flip dephasing from Eq.\eqref{eq:dephasing-expansion-main}. We choose the order $k$ based on the optimized noiseless QFI, as in Eq.\eqref{eq:noiseless-scaling}.}
    \label{fig:asymptotic-noise}
\end{figure}

\section{Exact QFI under iid dephasing noise}

We next apply an independent and identically distributed (iid) dephasing channel before encoding,
\begin{equation}
    \rho_p=\mathcal D_p^{\otimes n}(\rho),\qquad
    \mathcal D_p(\rho)=(1-p)\rho+pZ\rho Z ,
    \label{eq:dephasing-channel}
\end{equation}
with $0\leq p\leq1/2$. Complete hypergraph states remain permutationally invariant under the dephasing channel but do not necessarily lie within the symmetric Dicke state. Because the hypergraph unitary is diagonal and commutes with all $Z$ errors, the noisy state has the exact spectral decomposition
\begin{equation}
    \rho_p=\sum_{\bm z}p^{|\bm z|}(1-p)^{n-|\bm z|}\ket{H_{\bm z}}\bra{H_{\bm z}},\,\,\,\,\,
    \ket{H_{\bm z}}=Z^{\bm z}\ket{H} .
    \label{eq:dephasing-spectrum}
\end{equation}
The states under the dephasing channel $\ket{H_{\bm z}}$ are orthonormal, so the coefficients in Eq.~\eqref{eq:dephasing-spectrum} are the eigenvalues of $\rho_p$. 

The generators are sums of single-qubit operators, and permutation symmetry allows the squared matrix elements $\left|\langle a|J_\mu|b\rangle\right|^2$ in the mixed-state QFI formula, Eq.~\eqref{eq:qfi_mixed}, to be grouped according to the Hamming distance, which counts the single-qubit phase flips needed to change one eigenstate into the other. We define the Hamming distance phase kernel as
\begin{equation}
    \mathcal C_r+i\mathcal S_r=2^{-N}\sum_{w=0}^{N}K_w(r;N)e^{i\Delta_w},
    \label{eq:hamming-distance-kernels}
\end{equation}
where $N=n-1$ and the phase difference $\Delta_w=\Theta(w+1)-\Theta(w)$. The binary Krawtchouk polynomial in the combinatorial normalization is $ K_w(r;N)=\sum_{a=0}^{w}(-1)^a\binom ra\binom{N-r}{w-a},$ \cite{krawtchouk1929,levenshtein_krawtchouk_1995}. Here, $r$ counts the positions where the two phase-flip patterns differ, equivalently the Hamming distance, and $a$ counts how many of the $w$ ones lie in those positions.

The noise moments that contribute to the remaining mixed-state QFI are
\begin{equation}
    \begin{aligned}
        f_v&=v^2(1-2p)^2+4vpq,\\
        g_r&=\sum_{j=0}^{r}\binom rjR_{rj},\\
        h_r&=\sum_{j=0}^{r}\binom rj(r-2j)^2R_{rj}.
    \end{aligned}
    \label{eq:noise-moments}
\end{equation}
where $q=1-p$ and $ R_{rj}=\frac{(a-b)^2}{a+b},$ with $a=p^{r-j}q^j$ and $b=p^jq^{r-j}$. The factors $f_v$ and $g_r$ are adapted from the graph-state dephasing formulas by Shettell and Markham \cite{shettell_graph_2020}. The terms $\mathcal C_r$ and $\mathcal S_r$ define the phase structure of the hypergraph state, and $h_r$ is the additional second moment needed for complex phases.

The exact phase-flipped QFIs are
\begin{equation}
    \begin{aligned}
        F_Q[\rho_p,J_y]&=\frac12\sum_{r=0}^{n}\binom nr\left[\mathcal S_r^2 f_{n-r}g_r+\mathcal C_{r-1}^2h_r\right],\\
        F_Q[\rho_p,J_x]&=\frac12\sum_{r=0}^{n}\binom nr\left[\mathcal C_r^2 f_{n-r}g_r+\mathcal S_{r-1}^2h_r\right].
    \end{aligned}
    \label{eq:dephasing-fast}
\end{equation}
with $\mathcal C_{-1}=\mathcal S_{-1}=\mathcal S_n=\mathcal C_n=0$ whenever the terms are out of range. At the standard phase $\phi=\pi$, every $e^{i\Delta_w}$ is real and hence $\mathcal S_r=0$.

Alternatively, the QFI can also be evaluated by decomposing the density operator and collective generator into total-spin, or Schur-Weyl blocks \cite{chase_collective_2008,shammah_open_2018}. Nevertheless, evaluating the Krawtchouk polynomials and noise moments requires only $O(n^2)$ arithmetic. More importantly, this exact formula yields a fixed-rate asymptotic law.

For $\phi=\pi$, every fixed $k\geq3$, every fixed phase-flip rate $0\leq p<1/2$, and each $\mu\in\{x,y\}$, there exist finite constants $C,s$ and a rate $0\leq\Lambda_{k,p}<1$. With $\eta=1-2p$,
\begin{equation}
    \begin{aligned}
        F_Q(\rho_p,J_\mu)
        &=c_{p,\mu}^{(k)}n^2+d_{p,\mu}^{(k)}n+r_{p,\mu}^{(k)}(n),\\\\
        c_{p,\mu}^{(k)}&=\eta^2c_{0,\mu}^{(k)},\\
        d_{p,x}^{(k)}&=\eta^2a_k+4p(1-p)(1-b_k^2),\\
        d_{p,y}^{(k)}&=\eta^2(1-a_k)+4p(1-p)(1-b_k^2),
    \end{aligned}
    \label{eq:dephasing-expansion-main}
\end{equation}
with the remainder $|r_{p,\mu}^{(k)}(n)|\leq Cn^s\Lambda_{k,p}^{n}$. The dephasing Heisenberg coefficient is a factor $\eta^2$ of the noiseless Heisenberg coefficient $c_{0,\mu}^{(k)}$, which is strictly positive whenever $k\ge3$. Therefore, complete hypergraph states retain Heisenberg scaling after dephasing in $0\leq p<1/2$.

\section{Classical Fisher Information under both noises}

Quantum Fisher information (QFI) bounds the largest information that can be extracted over all suitable measurements, while the classical Fisher information (CFI) of a particular measurement does not necessarily saturate the QFI. Therefore, here we propose a local $Z$-measurement readout, or equivalently, Hamming-weight detection, to saturate the QFI at angle $\theta=0$ under $J_y$ encoding in the noiseless case. This follows from Theorem~2 of \cite{liuQuantumEstimationState2026}, since the noiseless encoded pure state given in Eq.~\eqref{eq:parameter_encoding} has only real amplitudes in the fixed computational basis. The corresponding classical Fisher information is
\begin{equation}
    F_C=\sum_{w=0}^{n-e}P_0(w)\left[\left.\partial_\theta\log P_\theta(w)\right|_{\theta=0}\right]^2,
    \label{eq:classical-cfi}
\end{equation}
where $P_\theta(w)$ is the probability of observing $w$ outcomes equal to $1$ among the $n-e$ surviving particles. For the complete hypergraph state at $\phi=\pi$ with $e$ particles erased and pre-encoding dephasing probability $p$, define the parity sign $s_w=(-1)^{\binom wk}$, the function $f(w)=s_ws_{w+1}=(-1)^{\binom w{k-1}}$, and the erasure-averaged function
\begin{equation}
    f_e(w)=2^{-e}\sum_{a=0}^{e}\binom ea f(w+a),
    \label{eq:cfi-average-erasures}
\end{equation}
For every $k\ge3$, $n\ge k$, and $m=n-e\ge2$, the CFI is
\begin{equation}
    \begin{aligned}
         F_C={}&\eta^2m\,\mathbb E[f_e(V_{m-1})^2]\\&+\frac{\eta^2m(m-1)}4\mathbb E\!\left[(\Delta f_e(V_{m-2}))^2\right],
    \end{aligned}
    \label{eq:local-cfi}
\end{equation}
where $\Delta f_e(V_{m-2})=f_e(V_{m-2}+1)-f_e(V_{m-2})$,$\eta=1-2p$ and $\mathbb{E}[\cdot]$ denotes averaging over the binomial random variable $V_t\sim\operatorname{Bin}(t,1/2)$. Let $P=2^{\lceil\log_2 k\rceil}$ be the period of $f$, and define
\begin{equation}
    \kappa_{e,k}=\frac1{4P}\sum_{w=0}^{P-1}[f_e(w+1)-f_e(w)]^2.
    \label{eq:cfi-positive-coefficient}
\end{equation}
For every fixed $k\ge3$, fixed finite $e$, and fixed $p<1/2$,
\begin{equation}
    \lim_{m\to\infty}\frac{F_C}{m^2}=\eta^2\kappa_{e,k}>0,
\end{equation}
shows that Heisenberg scaling is retained after local computational basis measurements. For $k=3$ and $m\ge3$, this simplifies into
\begin{equation}
    F_C=(1-2p)^2\frac{m(m+1)}{2^{e+1}}.
    \label{eq:cfi-k3}
\end{equation}

\begin{figure}
    \centering
    \includegraphics[width=\linewidth]{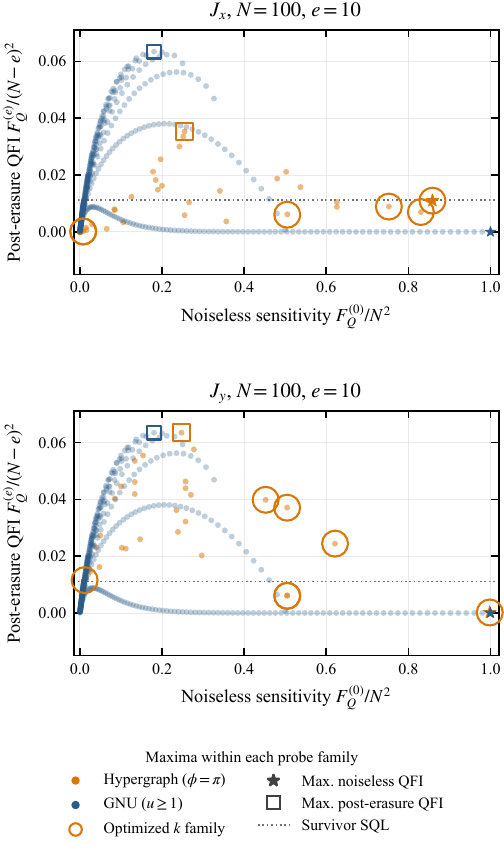}
    \caption{Noiseless sensitivity and post-erasure QFI for complete hypergraph families and GNU families by ref.\cite{ouyangRobustQuantumMetrology2022} under $J_x$ (Top) and $J_y$ (Bottom) collective sensing. We choose the optimized $k$ hypergraph families based on the optimized noiseless QFI, as in Eq.\eqref{eq:noiseless-scaling}.}
    \label{fig:gnu-hypergraph}
\end{figure}

\section{Discussion}
We show that complete weighted hypergraph states admit exact, polynomial-cost QFI equations \eqref{eq:qfi}, \eqref{eq:reduced_erasure_state}, and \eqref{eq:dephasing-fast} under collective generators $J_x$ and $J_y$. For the collective generator $J_z$, the symmetry of complete hypergraph states reduces the QFI to $n$, staying at the SQL. Nevertheless, one could still pre-rotate the state before the generator $J_z$ to the favorable axis. Fig.~\ref{fig:noiseless-envelopes} shows that complete hypergraph states achieve the HL asymptotically under $J_x$ but oscillate between the full and half HL under $J_y$ up to $n=1024$, where we optimize order $k$ at each $n$. 

For a finite-size system up to 100 qubits, we define a noise threshold as the maximum noise that the state can tolerate while keeping its post-noise QFI at the SQL. Fig. \ref{fig:noise-thresholds} demonstrates that complete hypergraph states are robust against particle erasure and iid dephasing noise. They are particularly robust under $J_y$ encoding, and a single order and a $\pi$ phase are sufficient to achieve the highest attainable robustness. While comparing with $J_x$ encoding, a complete weighted graph tends to have higher noise robustness, and sweeping the phase of higher-order hypergraph states increases the noise threshold. Therefore, it can be concluded that $J_y$ encodings are more noise-optimized. Higher erasure thresholds also yield lower noiseless QFI for the same state, showing erasure-sensitivity trade-offs; this is not true for iid dephasing thresholds, and one can find a higher noiseless QFI for a state with a high dephasing threshold. Moreover, one could also increase the number of orders in a complete hypergraph state, yielding more design choices but at the cost of preparing $\binom{n}{k}$ gates. Recent studies in \cite{zakaryan_nonsymmetric_2025} show that representing GHZ states using multi-order weighted complete hypergraphs could lead to HL QFI.

To assess robustness against particle erasure and (iid) phase flips before encoding, we require the Fisher information under noise to scale quadratically with the system size, which requires the corresponding Heisenberg coefficients to remain strictly positive as the system size grows. Equations \eqref{eq:erasure-expansion-main} and \eqref{eq:dephasing-expansion-main} establish this positivity under both types of noise, although the coefficients are not necessarily uniformly bounded away from zero. The coefficients decrease with increasing particle erasure and quadratically with the phase-flip parameter through the factor $(1-2p)^2$. Neglecting the remainders gives accurate finite-size approximations, although the error generally increases for higher $k$, $e$, and $p$. Unlike the noise threshold plot, Fig.\ref{fig:asymptotic-noise} shows that at states with equal or higher noiseless sensitivity, higher order $k$ yields a slower decay of the asymptotic Heisenberg coefficients and thus higher noise robustness without the sensitivity-erasure tradeoff. For higher erased particles $e$ and higher order $k$, convergence to the asymptotic Heisenberg coefficient limit generally requires a larger system size $n$.

It is also useful to compare the complete hypergraph states with a known robust resource, GNU states, which are QEC-inspired and represented in the Dicke basis. Fig.\ref{fig:gnu-hypergraph} shows that for an $N=100$ qubit system, after 10 particle erasures, high-sensitivity GNU states generally fall below the SQL limit, while complete hypergraph states provide access to the more favorable high-sensitivity and more erasure-robust region. For the $J_y$ generator, at a similar maximum post-erasure QFI, a hypergraph state configuration can achieve higher noiseless sensitivity than GNU states. 

Because the QFI is obtained by maximizing the classical Fisher information (CFI) over all possible measurements, attaining it generally requires an optimized measurement scheme \cite{braunstein_statistical_1994,toscano_attainability_2017}. For $\pi$-phase complete hypergraph states under $J_y$ encodings, we show that local computational-basis measurements saturate the noiseless state, suffice to attain Heisenberg scaling under erasure and dephasing noise, and can saturate the QFI in the noiseless case. Instead of recovering the $2^n$ strings, a Hamming weight detection needs only $n+1$ strings. For $J_x$ sensing, no simple optimal readout measurement generally saturates the noiseless QFI. Nevertheless, one can still obtain the optimal measurement from the eigenbasis of the symmetric logarithmic derivative (SLD) of the state, though it may not be local.

\vspace{\baselineskip}

\bibliography{references}

\clearpage
\setcounter{section}{0}
\setcounter{equation}{0}
\setcounter{figure}{0}
\setcounter{table}{0}
\renewcommand{\thesection}{S\arabic{section}}
\renewcommand{\theequation}{S\arabic{equation}}
\renewcommand{\thefigure}{S\arabic{figure}}
\renewcommand{\thetable}{S\arabic{table}}
\renewcommand{\theHsection}{S.\arabic{section}}
\renewcommand{\theHequation}{S.\arabic{equation}}
\renewcommand{\theHfigure}{S.\arabic{figure}}
\renewcommand{\theHtable}{S.\arabic{table}}

\section*{End Matter}
\label{sec:end_matter}

\section{Derivation of the noiseless QFI}
\label{app:derivation-noiseless}

We derive the noiseless QFI in Eq. \eqref{eq:qfi} by evaluating the one- and two-qubit correlators of the collective generators. For the state in Eq. \eqref{eq:state}, all computational-basis strings with the same Hamming weight have the same phase. For either $\mu=x$ or $y$, permutation invariance makes
\begin{equation}
    a_\mu=\langle\sigma_\mu^{(i)}\rangle,
    \qquad
    b_\mu=\langle\sigma_\mu^{(i)}\sigma_\mu^{(j)}\rangle
    \quad (i\neq j)
\end{equation}
independent of the chosen qubits. Since
$J_\mu=\frac12\sum_i\sigma_\mu^{(i)}$, we have
\begin{equation}
    \langle J_\mu\rangle=\frac n2a_\mu,
    \qquad
    \langle J_\mu^2\rangle=\frac14\left[n+n(n-1)b_\mu\right],
    \label{eq:app-collective-moments}
\end{equation}
and hence
\begin{equation}
    F_Q^{(\mu)}=4\,\mathrm{Var}(J_\mu)=n+n(n-1)b_\mu-n^2a_\mu^2.
    \label{eq:app-qfi-correlator}
\end{equation}

\subsection{One-body correlators}

Fix qubit $i$ and let $r$ be the number of ones among the other $n-1$ qubits. Flipping qubit $i$ modifies the Hamming weight of the string to
\begin{equation}
    \begin{array}{c|c|c|}
        x_i & |x| & |x\oplus e_i|\\ \hline
        0 & r   & r+1\\
        1 & r+1 & r\\
    \end{array}
    \label{eq:app-one-flip-cases}
\end{equation}
Their phase difference is $\Delta_r^{(1)} = \Theta(r+1)-\Theta(r) =\binom{r}{k-1}$ with Pascal's identity $\binom{r+1}{k} -\binom{r}{k} =\binom{r}{k-1}$.

For $X_i$, the two members of each pair contribute the conjugate phases $e^{-i\Delta_r^{(1)}}$ and $e^{i\Delta_r^{(1)}}$, that sum up into $2\cos\Delta_r^{(1)}$. For $Y_i$, the additional factors from $Y\ket0=i\ket1$ and $Y\ket1=-i\ket0$ give $i e^{-i\Delta_r^{(1)}}-i e^{i\Delta_r^{(1)}}=2\sin\Delta_r^{(1)}$.

There are $\binom{n-1}{r}$ configurations with a given $r$. After summing over $r$ and including the normalization $2^{-n}$, the resulting binomial averages are precisely those in Eq. \eqref{eq:noiseless-averages}:
\begin{equation}
    a_x=\langle X_i\rangle=C_1,
    \qquad
    a_y=\langle Y_i\rangle=S_1.
    \label{eq:app-one-body-correlators}
\end{equation}

\subsection{Two-body correlators}

Now fix distinct qubits $i$ and $j$, and let $u$ be the number of ones among the remaining $n-2$ qubits. Flipping both qubits gives
\begin{equation}
    \begin{array}{c|c|c}
        (x_i,x_j) & |x| & |x\oplus e_i\oplus e_j|\\ \hline
        (0,0) & u   & u+2\\
        (1,1) & u+2 & u\\
        (0,1),(1,0) & u+1 & u+1
    \end{array}
    \label{eq:app-two-flip-cases}
\end{equation}
Thus the $00$ and $11$ sectors acquire the phase difference $\Delta_u^{(2)} = \Theta(u+2)-\Theta(u) = 2\binom{u}{k-1} + \binom{u}{k-2}$ by applying the Pascal's identity twice. The $01$ and $10$ sectors retain the same Hamming weight and therefore have zero phase difference.

For $X_iX_j$, all four Pauli matrix elements have positive sign, so their combined contribution at fixed $u$ is
\begin{equation}
    e^{-i\Delta_u^{(2)}}+e^{i\Delta_u^{(2)}}+1+1=2+2\cos\Delta_u^{(2)} . 
\end{equation}

For $Y_iY_j$, the $00\leftrightarrow11$ matrix elements have a minus sign, whereas the $01\leftrightarrow10$ elements have a plus sign. Hence
\begin{equation}
    -e^{-i\Delta_u^{(2)}}-e^{i\Delta_u^{(2)}}+1+1=2-2\cos\Delta_u^{(2)} .
\end{equation}
Summing over the $\binom{n-2}{u}$ configurations shows that the two unchanged-weight sectors contribute $1/2$, while the remaining cosine average is $C_2$ from Eq. \eqref{eq:noiseless-averages}. Therefore
\begin{equation}
    b_x=\langle X_iX_j\rangle=\frac12+C_2,
    \qquad
    b_y=\langle Y_iY_j\rangle=\frac12-C_2.
    \label{eq:app-two-body-correlators}
\end{equation}

Finally, inserting Eqs.\eqref{eq:app-one-body-correlators} and \eqref{eq:app-two-body-correlators} into Eq.\eqref{eq:app-qfi-correlator} gives the exact $J_x$ and $J_y$ results in Eq.\eqref{eq:qfi} of the main text. For a single-order complete hypergraph state ($K=\{k\}$ and $\phi=\pi$), every $\Delta_r^{(1)}$ is an integer multiple of $\pi$, so $S_1=0$ and $b_y=1-b_x$.

\section{Exact large-\texorpdfstring{$n$}{n} limits of the single-order envelopes}
\label{app:noiseless-HC}

Throughout this section, $K=\{k\}$, $\phi=\pi$, and $3\leq k\leq n$.
At this phase, the trigonometric factors in
Eq.\eqref{eq:noiseless-averages} reduce exactly to binomial parities:
\begin{equation}
    \begin{aligned}
         \cos\Delta_r^{(1)}&=(-1)^{\binom{r}{k-1}},
         &\sin\Delta_r^{(1)}&=0,\\
         \cos\Delta_u^{(2)}&=(-1)^{\binom{u}{k-2}}.
    \end{aligned}
    \label{eq:app-parity-reduction}
\end{equation}
In the second line, the term $2\binom{u}{k-1}$ in
$\Delta_u^{(2)}/\pi$ is even and therefore does not affect the sign.

The binomial weights in Eq.\eqref{eq:noiseless-averages} are the
probabilities $\Pr(U_N=u)=2^{-N}\binom Nu$ for
$U_N\sim\operatorname{Bin}(N,1/2)$.  Thus the parity sums may be read as
expectations.  For any integer-valued random variable $B$,
\begin{equation}
    \mathbb E[(-1)^B]=1-2\Pr[B\ {\rm is\ odd}].
    \label{eq:binomial-mean}
\end{equation}
Lucas' theorem modulo two \cite{lucas_theorie_1878,fine_binomial_1947} gives
$\binom{U_N}{t}$ odd exactly when every one-bit of $t$ is also a one-bit
of $U_N$.  Let $s_2(t)$ be the number of one-bits of $t$ and define the
corresponding probability
\begin{equation}
    p_N(t)=\Pr\!\left[(U_N\mathbin{\&}t)=t\right].
    \label{eq:app-pnt}
\end{equation}
Here $\mathbin{\&}$ denotes bitwise AND. Then, the binomial sums in \eqref{eq:noiseless-averages} become
\begin{equation}
    \begin{aligned}
        C_1&=1-2p_{n-1}(k-1),
        &S_1&=0,\notag\\
        C_2&=\frac12-p_{n-2}(k-2).
        \label{eq:app-parity-averages}
    \end{aligned}    
\end{equation}

Substitution into the main-text QFI formula, Eq.\eqref{eq:qfi}, yields
\begin{equation}
    \begin{aligned}
        F_Q^{(x)}(n,k,\pi)
        &=n+n(n-1)\left[1-p_{n-2}(k-2)\right]\\
        &\quad-n^2\left[1-2p_{n-1}(k-1)\right]^2,\\
        F_Q^{(y)}(n,k,\pi)
        &=n+n(n-1)p_{n-2}(k-2).
        \label{eq:app-qfi-parity}
    \end{aligned}    
\end{equation}

For fixed $t$, choose $q=2^L>t$.  The classical Ramus identity for lacunary binomial sums, equivalently the roots-of-unity filter, shows that $U_N\bmod q$ approaches the uniform distribution exponentially \cite{RamusSolutionGenerale1834,konvalinaArithmeticProgressionSums1997}.  Exactly $2^{L-s_2(t)}$ of the $q$ residues contain all one-bits of $t$; hence
\begin{equation}
    \lim_{N\to\infty}p_N(t)=2^{-s_2(t)}.
    \label{eq:app-pnt-limit}
\end{equation}
Combining Equations \eqref{eq:app-qfi-parity}--\eqref{eq:app-pnt-limit} prove \eqref{eq:noiseless-fixed-coefficients}.  

The number of bits in a binary string $t$, denoted as $s_2(t)$, is 1 if $t=2^m$, $m$ if $t=2^m-1$, and 0 if $t=0$. To achieve the Heisenberg limit in $J_x$ encoding in \eqref{eq:noiseless-fixed-coefficients}, we require the $a_k$ and $b_k$ terms to be as small as possible; thus, when $k=2^m+1$, the $a_k$ term is $2^{-m}$, and the $b_k$ term is zero, producing equation \eqref{eq:noiseless-fixed-coefficients}. 

In $J_y$ encoding, we want to maximize $a_k$; $k=2$ obtains the HL, while $k=2^m+2$ reaches only half of the HL. The term that contributes to the Heisenberg scaling in Eq.\eqref{eq:app-qfi-parity} is $p_{n-2}(k-2)$, and to show that it approaches unity along a suitable sequence, choose $n=3\cdot2^m+2$ and $k=2^m+2$. Then, let $N=n-2=3\cdot2^m$ and $t=k-2=2^m=N/3$. Since $t$ has only one non-zero binary digit, the condition $(U_N\mathbin{\&}t)=t$ holds whenever
\begin{equation}
    \frac{N}{3}\leq U_N<\frac{2N}{3},
\end{equation}
with mean $\mathbb{E}[U_N]=N/2$. Then,
\begin{equation}
    \begin{aligned}
     p_N(t)
     &\geq \Pr\!\left[\frac{N}{3}\leq U_N<\frac{2N}{3}\right]\notag\\
     &\geq 1-\Pr\!\left[\left|U_N-\frac{N}{2}\right|
                       \geq\frac{N}{6}\right].
    \end{aligned}    
\end{equation}

Because $U_N$ is the sum of $N$ independent Bernoulli random variables with success probability $1/2$, Hoeffding's inequality \cite{hoeffding_probability_1963} gives
\begin{equation}
    \Pr\!\left[\left|U_N-\frac{N}{2}\right|\geq a\right]
    \leq 2\exp\!\left(-\frac{2a^2}{N}\right).
\end{equation}
With $a=N/6$ yields
\begin{equation}
    p_{n-2}(k-2)\geq 1-2e^{-(n-2)/18}.
\end{equation}
Thus $F_Q^{(y)}(n,k)/n^2\to1$ exponentially fast along the sequence $k=2^m+2$, $n=3\cdot2^m+2$ as $m\to\infty$. 

\section{Erasure Heisenberg Coefficient}
\label{app:erasure-HC}

Let $M=n-e$ be the number of particles remaining after $e$ erasures. Replacing $J_\mu$ by $J_\mu/M$ in Eq.~\eqref{eq:qfi_mixed} yields
\begin{equation}
    F_Q\left[\rho,\frac{J_\mu}{M}\right]=\frac{F_Q[\rho,J_\mu]}{M^2}.
\end{equation}
We therefore study the generator $J_\mu/M$ to find the quadratic-term coefficient.

Using the Fourier branches defined above, the erased state is
\begin{equation}
    \rho_M^{(e)}=\sum_{q,q'}(\sigma_e)_{qq'}\ket{\theta_q}^{\otimes M}\bra{\theta_{q'}}^{\otimes M},
\end{equation}
where
\begin{equation}
    (\sigma_e)_{qq'}=a_qa_{q'}^*\left[\frac{1+e^{i(\theta_q-\theta_{q'})}}{2}\right]^e.
    \label{eq:HC-branch-state}
\end{equation}
Let $\Pi_M$ be the orthogonal projector onto the span of the active branches,
\begin{equation}
    \mathcal S_M=\operatorname{span}\left\{\ket{\theta_q}^{\otimes M}:a_q\neq0\right\}.
\end{equation}
Since $\rho_M^{(e)}=\Pi_M\rho_M^{(e)}\Pi_M$, the QFI separates exactly into
\begin{equation}
    \begin{aligned}
        F_Q[\rho_M^{(e)},J_\mu]={}&F_Q[\rho_M^{(e)},\Pi_MJ_\mu\Pi_M]\\&+4\operatorname{Tr}\left[\rho_M^{(e)}J_\mu(I\Pi_M)J_\mu\right].
        \label{eq:HC-QFI-split}
    \end{aligned}    
\end{equation}

The first term describes the generator acting within the branch span. The second describes its action outside this span. They produce the quadratic and linear terms, respectively.

For fixed $k$, the overlaps between different Fourier branches decrease exponentially with $M$:
\begin{equation}
    \left|\langle\theta_q|\theta_{q'}\rangle\right|^M=\left|\cos\frac{\theta_q-\theta_{q'}}{2}\right|^M\leq\gamma_k^M
\end{equation}
for $\gamma_k<1$ and $\quad q\neq q'$.Thus the branches become orthogonal as $M$ increases, and the state within the branch span is represented by $\sigma_e$, while $J_\mu/M$ is represented by the mean spin per particle on each branch:
\begin{equation}
    (h_x)_{qq'}=\frac{\cos\theta_q}{2}\delta_{qq'},
    \qquad
    (h_y)_{qq'}=\frac{\sin\theta_q}{2}\delta_{qq'}.
\end{equation}
The quadratic coefficient is therefore
\begin{equation}
    c_{e,\mu}^{(k)} = \lim_{M\to\infty} \frac{ F_Q[\rho_M^{(e)},\Pi_MJ_\mu\Pi_M] }{M^2}=F_Q[\sigma_e,h_\mu].
\label{eq:HC-quadratic}
\end{equation}
This coefficient can be evaluated directly using Eq.~\eqref{eq:qfi_mixed}, and the size of the matrices only depends on $k$.

Next, the linear term comes from the spin fluctuations within each product branch under the second term in Eq.\eqref{eq:HC-QFI-split}. Since different particles in a product state have no correlations, their variances sum up:
\begin{equation}
    \operatorname{Var}_{\theta_q^{\otimes M}}(J_\mu)=\frac{M}{4}\left[1-4h_\mu(q)^2\right],
\end{equation}
As the branches become orthogonal, the second term in Eq.~\eqref{eq:HC-QFI-split} is four times the average of these variances, up to exponentially small corrections. Since $(\sigma_e)_{qq}=|a_q|^2$, its linear coefficient is
\begin{equation}
    d_\mu^{(k)}=\sum_q|a_q|^2\left[1-4h_\mu(q)^2\right].
    \label{eq:HC-linear}
\end{equation}
This coefficient is independent of $e$ because erasure leaves the diagonal branch weights $|a_q|^2$ unchanged.

Using Lucas's Theorem and the roots-of-unity filter, the Fourier terms satisfy
\begin{equation}
    \sum_q|a_q|^2\cos(2\theta_q)=1-2^{1-s_2(k-2)},
\end{equation}
where $s_2(t)$ is the number of the ones in binary $t$. Using $\sum_q|a_q|^2=1$ and the identities for $\sin^2\theta$ and $\cos^2\theta$ gives
\begin{equation}
    d_x^{(k)}=2^{-s_2(k-2)},
    \qquad
    d_y^{(k)}=1-2^{-s_2(k-2)}.
\end{equation}

Combining the two contributions yields the equation \eqref{eq:erasure-expansion-main}. Next, $F_Q[\sigma_e,h_\mu]>0$ whenever $[\sigma_e,h_\mu]\neq0$. Since $h_\mu$ is diagonal,
\begin{equation}
    [\sigma_e,h_\mu]_{qq'}=(\sigma_e)_{qq'}\left[h_\mu(q')-h_\mu(q)\right].
\end{equation}
For $L\ge8$, or equivalently $k\ge4$, the $(q,q')$ pairs of $(1, 3)$ for $J_x$ and $(1, L-1)$ for $J_y$ have different generator values, and thus the non-diagonal terms remain nonzero after every fixed finite $e$. For $k=3$, the pair $(0, 1)$ has different values. Therefore, these prove the strict positivity of the erasure Heisenberg coefficient term
\begin{equation}
    \lim_{M\to\infty}\frac{F_Q[\rho_M^{(e)},J_\mu]}{M^2}=c_{e,\mu}^{(k)}>0,
\end{equation}
which retain Heisenberg scaling under erasure noise.

\section{Derivation of the QFI under phase flips}
\label{app:derrivation-dephasing}

We insert the spectral decomposition in Eq.~\eqref{eq:dephasing-spectrum} into the mixed-state QFI formula, Eq.~\eqref{eq:qfi_mixed}. Let $p$ be the phase-flip probability and $q=1-p$. For two error strings $(\bm z,\bm z')$, define their disagreement set $D=\operatorname{supp}(\bm z\oplus\bm z')$, with Hamming distance $r=|D|$, $v=n-r$ and $\sigma_i=(-1)^{z_i}$. The disagreement set $D$ contains the $r$ positions where the patterns differ, while $D^{\mathrm c}$ contains the $v$ positions where they agree.

Next, define
\begin{equation}
    j=\sum_{i\in D}z_i,\\\\
    \ell=\sum_{i\notin D}z_i,
\end{equation}
as the number of errors in the disagreement and common positions, respectively. Since $\bm z'$ agrees with $\bm z$ outside $D$ and is its complement on $D$, the eigenvalues are
\begin{equation}
\begin{aligned}
    \lambda_{\bm z}&=p^\ell q^{v-\ell}p^jq^{r-j},\\
    \lambda_{\bm z'}&=p^\ell q^{v-\ell}p^{r-j}q^j.
\end{aligned}
\label{eq:app-dephasing-eigenvalue-factorization}
\end{equation}
Consequently, the eigenvalue factor in the QFI becomes
\begin{equation}
    \frac{(\lambda_{\bm z}-\lambda_{\bm z'})^2}{\lambda_{\bm z}+\lambda_{\bm z'}}=p^\ell q^{v-\ell}R_{rj}(p).
\label{eq:app-dephasing-spectral-ratio}
\end{equation}

Next, fix a qubit $i$ on which $X_i$ or $Y_i$ acts. Define $\bm u\in\{0,1\}^{N}$ for the configuration of the other $N=n-1$ bits and $w=|\bm u|$ for their Hamming weight. The two configurations with $x_i=0$ and $x_i=1$ have total weights $w$ and $w+1$. Flipping qubit $i$ therefore introduces the phase difference $\Delta_w=\Theta(w+1)-\Theta(w)$.

For fixed $\bm u$, pairing the two values of $x_i$ gives the factors
\begin{equation}
    \begin{aligned}
        X_i:\quad&\frac12\left[\sigma_i e^{-i\Delta_w}+(-1)^{z_i'}e^{i\Delta_w}\right],\\
        Y_i:\quad&\frac{i}{2}\left[\sigma_i e^{-i\Delta_w}-(-1)^{z_i'}e^{i\Delta_w}\right].
    \end{aligned}
\end{equation}

The second line follows from $Y\ket0=i\ket1$ and $Y\ket1=-i\ket0$. Both factors are multiplied by the remaining-bit error sign $(-1)^{|\bm u_{D_i}|}$, where
\begin{equation}
    D_i=D\setminus\{i\},
    \qquad
    s=|D_i|=
    \begin{cases}
        r,&i\notin D,\\
        r-1,&i\in D.
    \end{cases}
\end{equation}

Here $|\bm u_{D_i}|$ counts the ones of $\bm u$ at positions in $D_i$. The required average over the remaining bits is
\begin{equation}
    \mathcal C_s+i\mathcal S_s=2^{-N}\sum_{\bm u\in\{0,1\}^{N}}(-1)^{|\bm u_{D_i}|}e^{i\Delta_{|\bm u|}}.
\label{eq:app-dephasing-radial-kernels}
\end{equation}
For total weight $w$, configurations with $a$ ones in $D_i$ have multiplicity $\binom sa\binom{N-s}{w-a}$ and sign $(-1)^a$. Grouping by $w$ therefore gives the Krawtchouk transform in Eq.~\eqref{eq:hamming-distance-kernels}.

When $i\notin D$, the two error signs agree, and the paired factors are $\sigma_i\cos\Delta_w$ for $X_i$ and
$\sigma_i\sin\Delta_w$ for $Y_i$. When $i\in D$, the signs are opposite, giving $-i\sigma_i\sin\Delta_w$ for $X_i$ and $i\sigma_i\cos\Delta_w$ for $Y_i$. Averaging over $\bm u$ and summing over all generator sites gives
\begin{equation}
    \begin{aligned}
        \bra{H_{\bm z}}\sum_iX_i\ket{H_{\bm z'}}&=\mathcal C_r A-i\mathcal S_{r-1}B,\\
        \bra{H_{\bm z}}\sum_iY_i\ket{H_{\bm z'}}&=\mathcal S_r A+i\mathcal C_{r-1}B,
    \end{aligned}
    \label{eq:app-dephasing-generator-elements}
\end{equation}
where
\[
A=\sum_{i\notin D}\sigma_i=v-2\ell,
\qquad
B=\sum_{i\in D}\sigma_i=r-2j.
\]
These are scalar sign sums. The local Pauli action is already included in $\mathcal C_s$ and $\mathcal S_s$, while $A$ and $B$ collect the contributions from the different generator sites. Because the two terms in each matrix element are real and purely imaginary, respectively, their squared modulus has no cross term.

We now sum over the error patterns. On the common positions, $p^\ell q^{v-\ell}$ is the probability of each pattern, so
\begin{equation}
    \begin{aligned}
         f_v(p)
         &=\sum_{\bm z_{D^{\mathrm c}}}
         p^\ell q^{v-\ell}A^2\\
         &=\mathbb E\!\left[
         \left(\sum_{i\notin D}\sigma_i\right)^2
         \right]\\
         &=\sum_{i\notin D}\mathbb E[\sigma_i^2]
        +\sum_{\substack{i,j\notin D\ i\neq j}}
        \mathbb E[\sigma_i\sigma_j]\\
         &=v+v(v-1)(1-2p)^2.
    \end{aligned}
    \label{eq:app-dephasing-f-derivation}
\end{equation}

On the disagreement positions, there are $\binom rj$ patterns with $j$ errors, and $B=r-2j$. Hence
\begin{equation}
    \begin{aligned}
         g_r(p)
         &=\sum_{\bm z_D}R_{rj}(p)
         =\sum_{j=0}^{r}\binom rjR_{rj}(p),\\
         h_r(p)
         &=\sum_{\bm z_D}R_{rj}(p)B^2\\
         &=\sum_{j=0}^{r}\binom rj(r-2j)^2R_{rj}(p).
    \end{aligned}
    \label{eq:app-dephasing-gh-derivation}
\end{equation}

For a fixed disagreement set $D$, choosing $\bm z$ uniquely determines $\bm z'$. The complete error-weighted sums therefore factorize as
\begin{equation}
    \begin{aligned}
        &\sum_{\bm z}p^\ell q^{v-\ell}R_{rj}A^2\\
        &\quad=
        \left(\sum_{\bm z_{D^{\mathrm c}}}
        p^\ell q^{v-\ell}A^2\right)
        \left(\sum_{\bm z_D}R_{rj}\right)
        =f_vg_r,\\[3pt]
        &\sum_{\bm z}p^\ell q^{v-\ell}R_{rj}B^2\\
        &\quad=
        \left(\sum_{\bm z_{D^{\mathrm c}}}
        p^\ell q^{v-\ell}\right)
        \left(\sum_{\bm z_D}R_{rj}B^2\right)
        =h_r.
    \end{aligned}
    \label{eq:app-dephasing-error-sums}
\end{equation}
In the second expression, the common-position probabilities sum to $(p+q)^v=1$. Thus the $A^2$ contribution produces $f_vg_r$, whereas the $B^2$ contribution produces $h_r$.

Finally, there are $\binom nr$ disagreement sets of size $r$. Summing their contributions in Eq.~\eqref{eq:qfi_mixed} gives the two QFI expressions in Eq.~\eqref{eq:dephasing-fast}. Their prefactor $1/2$ is the product of the factor $2$ in the spectral QFI formula and the factor $1/4$ from squaring $J_\mu=\frac12\sum_i\sigma_\mu^{(i)}$.

\section{Dephasing Heisenberg Coefficient}
\label{app:deph-HC}
Here, we consider the complete $k$-uniform hypergraph state with $\phi=\pi$ and fixed $k\geq3$ under identical and independent Pauli-$Z$ flips before parameter encoding. Let $q=1-p$, $\eta=1-2p$, with $0<p<1/2$. Rewrite the $R_{rj}$ terms inside the noise moments in Eq.~\eqref{eq:noise-moments} as $R_{rj}=a+b-4ab/(a+b)$, with $a=p^{r-j}q^j$ and $b=p^jq^{r-j}$. The binomial sum of $a+b$ is $2$, while its $(r-2j)^2$-weighted sum is $2(\eta^2r^2+4pqr)$. The arithmetic-geometric mean inequality,
\begin{equation}
    \frac{a+b}{2}\geq\sqrt{ab},
    \label{AM-GM-inequality}
\end{equation}
gives $4ab/(a+b)\leq2(pq)^{r/2}$. Since $\sum_j\binom rj(r-2j)^2=r2^r$, the noise moments can be written as
\begin{equation}
    \begin{aligned}
        g_r&=2-\varepsilon_r,
        &0\leq\varepsilon_r&\leq2\alpha^r,
        \notag\\
        h_r&=2(\eta^2r^2+4pqr)-\delta_r,
        &0\leq\delta_r&\leq2r\alpha^r,
        \label{eq:dhc-spectral-bounds}
    \end{aligned}    
\end{equation}
for $r\geq1$, where $\alpha=2\sqrt{pq}<1$. Note that $g_0=0$ and $\delta_1=8pq$, which will later be considered in the coefficient formula.

At $\phi=\pi$, the QFIs are
\begin{equation}
    \begin{aligned}
         F_Q(\rho_p,J_x)
         &=\frac12\sum_{r=0}^{n-1}
         \binom nr c_r^2f_{n-r}g_r,
         \label{eq:dhc-exact-x}\\
         F_Q(\rho_p,J_y)
         &=\frac12\sum_{r=1}^{n}
         \binom nr c_{r-1}^2h_r,
    \end{aligned}    
\end{equation}
where $f_v=\eta^2v^2+4pqv$. Put $N=n-1$ and $\sigma_w=(-1)^{\binom{w}{k-1}}$, the coefficient
\begin{equation}
    \begin{aligned}
        c_r&=2^{-N}\sum_{w=0}^{N}K_w(r;N)\sigma_w,\\
        K_w(r;N)&=\sum_a(-1)^a
        \binom ra\binom{N-r}{w-a}.
    \end{aligned}
    \label{eq:dhc-radial}
\end{equation}

The sums over $r$ simplify through Krawtchouk orthogonality,
\begin{equation}
    \sum_{r=0}^{N}\binom Nr K_w(r;N)K_u(r;N)=2^N\binom Nw\delta_{wu}.
    \label{eq:dhc-orthogonality}
\end{equation}

Substituting Eq.~\eqref{eq:dhc-radial} and using $\sigma_w^2=1$ gives the normalization
\begin{equation}
    \sum_{r=0}^{N}\binom Nr c_r^2=2^{-N}\sum_{w=0}^{N}\binom Nw\sigma_w^2=1.
    \label{eq:dhc-cr2}
\end{equation}

Using the orthogonality condition and the Krawtchouk recurrence formula
\begin{equation}
    (N-2r)K_w=(w+1)K_{w+1}+(N-w+1)K_{w-1}.
    \label{eq:dhc-recurrence}
\end{equation}
yields
\begin{equation}
    \sum_{r=0}^{N}r\binom Nr c_r^2
    =\frac N2-\frac{N}{2^N}
    \sum_{w=0}^{N-1}\binom{N-1}{w}\sigma_w\sigma_{w+1}.
    \label{eq:dhc-weighted-moment}
\end{equation}
Pascal identity then gives $\sigma_w\sigma_{w+1} = (-1)^{\binom{w}{k-2}}$.

Using Lucas' theorem and the same probability definition in Eq\eqref{eq:app-pnt} yields
\begin{equation}
    \begin{aligned}
         \sum_{r=0}^{N}r\binom Nr c_r^2
         &=\frac N2-\frac N2[1-2p_{N-1}(k-2)]
         \\
         &=Np_{N-1}(k-2).
     \label{eq:dhc-rcr2}
    \end{aligned}
\end{equation}
The coefficient $c_0$ is another binomial parity average with
\begin{equation}
c_0=1-2p_{n-1}(k-1).
\label{eq:dhc-c0}
\end{equation}
In $J_x$ encoding, use $N=n-1$ and substitute $g_r=2-\varepsilon_r$ for $r\geq1$ and $g_0=0$ to get
\begin{equation}
    F_Q(\rho_p,J_x)=\sum_{r=0}^{n-1}\binom nr c_r^2f_{n-r}-c_0^2f_n-C_x,
\end{equation}
where
\begin{equation}
    C_x:=\frac12\sum_{r=1}^{n-1}
    \binom nr c_r^2f_{n-r}\varepsilon_r\geq0.
    \label{eq:dhc-correction-x}
\end{equation}
Using $(n-r)\binom nr=n\binom{n-1}{r}$ together with Eq.\eqref{eq:dhc-cr2}, \eqref{eq:dhc-rcr2} and \eqref{eq:dhc-c0} yields
\begin{equation}
    \begin{aligned}
        F_Q(\rho_p,J_x)
        ={}&\eta^2n\bigl[n-(n-1)p_{n-2}(k-2)\bigr]
        \notag\\
        &-\eta^2n^2c_0^2+4pqn(1-c_0^2)-C_x.
        \label{eq:dhc-x-finite}
    \end{aligned}
\end{equation}
In $J_y$ encoding, using identity $r\binom nr=n\binom{n-1}{r-1}$ gives
\begin{equation}
    \begin{aligned}
         \sum_{r=1}^{n}r\binom nr c_{r-1}^2&=n,
         \\
         \sum_{r=1}^{n}r^2\binom nr c_{r-1}^2&=n\bigl[1+(n-1)p_{n-2}(k-2)\bigr],
    \end{aligned}
\end{equation}
and for $r=1$, the term contributes $\frac12nc_0^2\delta_1=4pqnc_0^2$. Including it yields 
\begin{equation}
    \begin{aligned}
         F_Q(\rho_p,J_y)={}&\eta^2n\bigl[1+(n-1)p_{n-2}(k-2)\bigr]
         \\
         &+4pqn(1-c_0^2)-C_y,
     \label{eq:dhc-y-finite}
    \end{aligned}
\end{equation}
where
\begin{equation}
    C_y:=\frac12\sum_{r=2}^{n}
    \binom nr c_{r-1}^2\delta_r\geq0.
    \label{eq:dhc-correction-y}
\end{equation}
Both Eqs.~\eqref{eq:dhc-x-finite} and \eqref{eq:dhc-y-finite} are exact and the $C_x$ and $C_y$ terms decays exponentially with higher $r$. Finally, using the roots-of-unity filter in Eq.\eqref{eq:app-pnt-limit} in the large-$n$ limit, we prove the formula in Eq.\eqref{eq:dephasing-expansion-main} of the main text.

\section{Derivation of CFI under both noises}
\label{app:cfi-both-noises}
For $\pi$-phase single-order $k$-uniform complete hypergraphs, Eq.~\eqref{eq:state} reduces to
\begin{equation}
    \begin{aligned}
        \ket{H_n(\pi;k)}&=2^{-n/2}\sum_x s_{|x|}\ket{x},\\
        s_w&=(-1)^{\binom wk}.
    \end{aligned}
    \label{eq:app-cfi-state}
\end{equation}
Since erasure noise and independent local dephasing can be applied in either order, we define the mixed state after both noises as $\rho_{e,p}=\mathcal D_p^{\otimes m}(\rho_m^{(e)})$. The computational-basis matrix elements of the resulting state are
\begin{equation}
    \begin{aligned}
        \langle y|\rho_{e,p}|z\rangle
        &=\frac{\eta^{d_H(y,z)}}{2^{m+e}}
        \sum_{t=0}^{e}\binom{e}{t}
        s_{|y|+t}s_{|z|+t},
    \end{aligned}
\label{eq:app-cfi-matrix-elements}
\end{equation}
with $y,z\in\{0,1\}^{m}$, $t\in\{0,\ldots,e\}$, $n=m+e$, $\eta=1-2p$, and Hamming distance $d_H(y,z)$. The diagonal entries, where $y=z$, give
$\langle y|\rho_{e,p}|y\rangle=2^{-m}$. Define the erasure average function by
\begin{equation}
    f_e(w)=2^{-e}\sum_{t=0}^{e}\binom{e}{t}f(w+t),
    \label{erasure-average}
\end{equation}
where $f(w)=s_ws_{w+1}=(-1)^{\binom{w}{k-1}}$. If $y^{(i)}$ is obtained from $y$ by flipping bit $i$, then Eq.~\eqref{eq:app-cfi-matrix-elements} becomes
\begin{equation}
    \langle y|\rho_{e,p}|y^{(i)}\rangle
    =\eta\,2^{-m}f_e(|y|-y_i).
    \label{eq:app-cfi-neighbor-coherence}
\end{equation}
Since $Y|0\rangle=i|1\rangle$ and $Y|1\rangle=-i|0\rangle$, the only nonzero entries in row $y$ of $J_y=\frac12\sum_{i=1}^{m}Y_i$ connect $y$ to $y^{(i)}$:
\begin{equation}
    \begin{aligned}
        (J_y)_{y,y^{(i)}}&=-\frac{i}{2}(-1)^{y_i},\\
        (J_y)_{y^{(i)},y}&=\frac{i}{2}(-1)^{y_i}.
    \end{aligned}
    \label{eq:app-cfi-generator-entries}
\end{equation}
The encoded state is
\begin{equation}
    \rho_\theta=e^{-i\theta J_y}\rho_{e,p}e^{i\theta J_y},
    \label{eq:app-cfi-encoding}
\end{equation}
and the full-string probability for local $Z$ measurements is $q_\theta(y)=\langle y|\rho_\theta|y\rangle$. Then, its derivative at $\theta=0$ is
\begin{equation}
    \begin{aligned}
        \dot q_0(y)
        &=\langle y|\dot\rho_0|y\rangle\\
        &=\langle y|-i[J_y,\rho_{e,p}]|y\rangle\\
        &=-\sum_{i=1}^{m}(-1)^{y_i}
        \operatorname{Re}\langle y|\rho_{e,p}|y^{(i)}\rangle\\
        &=-\eta\,2^{-m}\sum_{i=1}^{m}
        (-1)^{y_i}f_e(|y|-y_i).
    \end{aligned}
\end{equation}
Using Hamming weight $w=|y|$ and grouping the terms into $m-w$ zero bits and $w$ one bits, the equation reduces to
\begin{equation}
    \dot q_0(y)
    =-\eta\,2^{-m}
    \bigl[(m-w)f_e(w)-wf_e(w-1)\bigr].
\end{equation}
At $w=0$, the term $wf_e(w-1)$ is omitted. The probability of detecting $w$ outcomes equal to one is
\begin{equation}
    \begin{aligned}
        P_\theta(w)&=\sum_{|y|=w}q_\theta(y),\\
        P_0(w)&=2^{-m}\binom{m}{w}.
    \end{aligned}
    \label{eq:app-cfi-weight-probability}
\end{equation}
Since $\dot q_0(y)$ depends only on $w$, $\dot P_0(w)=\binom{m}{w}\dot q_0(y)$ for $|y|=w$. Then, the CFI at $\theta=0$ is
\begin{equation}
    \begin{aligned}
        F_C
        &=\sum_y\frac{\dot q_0(y)^2}{q_0(y)}\\
        &=\sum_{w=0}^{m}\frac{\dot P_0(w)^2}{P_0(w)}\\
        &=\eta^2\,2^{-m}\sum_{w=0}^{m}\binom{m}{w}\\
        &\qquad\times
        \bigl[(m-w)f_e(w)-wf_e(w-1)\bigr]^2.
    \end{aligned}
\end{equation}
For $m\ge2$, let $V_r\sim\operatorname{Bin}(r,\tfrac12)$.Expanding the square, using $w^2=w(w-1)+w$ gives
\begin{equation}
    \begin{aligned}
        F_C={}&\eta^2m\,\mathbb E[f_e(V_{m-1})^2]\\
        &+\frac{\eta^2m(m-1)}4\,
        \mathbb E\Bigl[
        f_e(V_{m-2})^2\\
        &\qquad+f_e(V_{m-2}+1)^2\\
        &\qquad-2f_e(V_{m-2})f_e(V_{m-2}+1)
        \Bigr].
    \end{aligned}
\end{equation}
The second expectation is a squared difference; simplifying it then proves Eq.\eqref{eq:local-cfi}.

\subsection{Exact expression for complete three-uniform }
For $k=3$, the sequence $f$ repeats $(1,1,-1,-1)$, so
\begin{equation}
    f(w)=\sqrt2\cos\!\left(\frac{\pi w}{2}-\frac\pi4\right).
\end{equation}
Using $\mathbb E[i^{V_e}]=[(1+i)/2]^e$ in the erasure average formula gives
\begin{equation}
    f_e(w)=2^{(1-e)/2}\cos\!\left(\frac{\pi w}{2}+\frac{(e-1)\pi}{4}\right).
\end{equation}
Then,
\begin{equation}
    \begin{aligned}
        f_e(w)^2&=2^{-e}\left[1+(-1)^w c_e\right],\\
        [\Delta f_e(w)]^2
        &=2^{1-e}\left[1+(-1)^w d_e\right],
    \end{aligned} 
\end{equation}
with $c_e=\cos\!\left(\frac{(e-1)\pi}{2}\right)$ and $d_e=\sin\!\left(\frac{(e-1)\pi}{2}\right)$.For every positive integer $t$, $\mathbb E[(-1)^{V_t}]=0$:
\begin{equation}
    \begin{aligned}
        \mathbb E[f_e(V_{m-1})^2]&=2^{-e},\\
        \mathbb E\bigl[(\Delta f_e(V_{m-2}))^2\bigr]&=2^{1-e},
    \end{aligned}
\end{equation}
which then yields
\begin{equation}
    F_C=(1-2p)^2\frac{m(m+1)}{2^{e+1}},
    \label{eq:app-cfi-k3}
\end{equation}
for $k=3$ and $m\ge 3$.

\subsection{CFI Heisenberg Coefficient}
The first term in Eq.\eqref{eq:local-cfi} is at most linear in $m$, whereas the second may scale as $m^2$. Therefore, to prove Heisenberg scaling, it is sufficient to determine whether the term
\begin{equation}
    \mathbb E\!\left[
    \bigl(\Delta f_e(V_{m-2})\bigr)^2
    \right]
\end{equation}
approaches a nonzero constant as $m\to\infty$. By Lucas' Theorem \cite{fine_binomial_1947}, the parity of $\binom{w}{k-1}$ depends only on a finite number of binary digits of $w$. Threfore, $f(w)$ is periodic with some finite period $P$ with
\begin{equation}
    f(w)=\sum_{\ell=0}^{P-1}\widehat f_\ell e^{i\omega_\ell w},
    \qquad
    \omega_\ell=\frac{2\pi\ell}{P}.
\end{equation}
For fixed $P$, the roots-of-unity-filter gives
\begin{equation}
    \mathbb E[h(V_s)]
    \longrightarrow
    \frac1P\sum_{w=0}^{P-1}h(w).
\end{equation}
Substituting this to the quadratic term in Eq.\eqref{eq:local-cfi} yields
\begin{equation}
    \kappa_{e,k}=\frac1{4P}\sum_{w=0}^{P-1}[f_e(w+1)-f_e(w)]^2.
    \label{eq:app-kappa-definition}
\end{equation}
The erasure average then becomes
\begin{equation}
    \begin{aligned}
    f_e(w)
    &=\sum_{\ell}\widehat f_\ell\omega^{\ell w}\left[2^{-e}\sum_{a=0}^{e}\binom ea\omega^{\ell a}\right]\\
    &=\sum_{\ell}\widehat f_\ell\left(\frac{1+\omega^\ell}{2}\right)^e\omega^{\ell w}.
    \end{aligned}
    \label{eq:app-fe-fourier}
\end{equation}
Hence mode $\ell$ is attenuated by
\begin{equation}
    \left|\frac{1+\omega^\ell}{2}\right|^{2e}=\cos^{2e}\left(\frac{\pi\ell}{P}\right).
\end{equation}
Define
\begin{equation}
    \mathcal A_e=
    \frac1P\sum_{w=0}^{P-1}f_e(w)^2,
    \qquad
    \kappa_e=
    \frac1{4P}\sum_{w=0}^{P-1}
    [\Delta f_e(w)]^2.
\end{equation}
Fourier orthogonality gives
\begin{equation}
    \mathcal A_e=
    \sum_\ell
    |\widehat f_\ell|^2
    \cos^{2e}\left(\frac{\pi\ell}{P}\right),
\end{equation}
and
\begin{equation}
    \kappa_e=
    \sum_{\ell=0}^{P-1}
    |\widehat f_\ell|^2
    \cos^{2e}\left(\frac{\pi\ell}{P}\right)
    \sin^2\left(\frac{\pi\ell}{P}\right),
    \label{eq:app-kappa}
\end{equation}
where we used $|e^{2\pi i\ell/P}-1|^2=4\sin^2(\pi\ell/P)$.
Every term is nonnegative. For $k\ge3$,
\begin{equation}
    f(0)=f(1)=1,
    \qquad
    f(k-1)=-1,
\end{equation}
so $f$ cannot consist only of the constant and parity Fourier modes. Hence at least one nonzero coefficient satisfies $\ell\neq0, P/2$. For this mode,
\begin{equation}
    \cos^{2e}\!\left(\frac{\pi\ell}{P}\right)>0,
    \qquad
    \sin^2\!\left(\frac{\pi\ell}{P}\right)>0
\end{equation}
for every finite $e$. Therefore, $\kappa_{e,k}>0$.

\end{document}